\documentclass[a4paper,10pt]{article}

\pdfoutput=1
\usepackage[utf8]{inputenc}
\usepackage{fullpage}
\usepackage{amsmath}
\usepackage{amsfonts}
\usepackage{amssymb}
\usepackage[pdftex]{hyperref}
\usepackage{graphicx}
\usepackage{array}
\usepackage{subfigure}
\usepackage{algorithmic}
\usepackage{algorithm}
\usepackage[numbers,sort&compress]{natbib}
\usepackage{authblk}
\usepackage{flushend}

\title{Position paper: a general framework for applying machine learning techniques in operating room}

\author[1*]{Filippo Maria Bianchi\thanks{filippo.binachi@ryerson.ca}\thanks{Corresponding author}}
\author[1*]{Enrico De Santis\thanks{enrico.desantis@ryerson.ca}}
\author{Hedieh Montazeri\thanks{hedieh.montazeri@ryerson.ca}}
\author{Parisa Naraei \thanks{parisa.naraei@ryerson.ca}}
\author{Alireza Sadeghian \thanks{asadeghi@ryerson.ca}}
\affil[]{Dept. of Computer Science, Ryerson University, 350 Victoria Street, Toronto, ON M5B 2K3, Canada}
\affil[*]{\textit{These two authors contributed equally}}

\begin{document}

\maketitle

\begin{abstract}
In this position paper we describe a general framework for applying machine learning and pattern recognition techniques in healthcare. In particular, we are interested in providing an automated tool for monitoring and incrementing the level of awareness in the operating room and for identifying human errors which occur during the laparoscopy surgical operation.
The framework that we present is divided in three different layers: each layer implements algorithms which have an increasing level of complexity and which perform functionality with an higher degree of abstraction. In the first layer, raw data collected from sensors in the operating room during surgical operation, they are pre-processed and aggregated. The results of this initial phase are transferred to a second layer, which implements pattern recognition techniques and extract relevant features from the data. Finally, in the last layer, expert systems are employed to take high level decisions, which represent the final output of the system.
\end{abstract}



\section{Introduction}
Operating room has always received considerable attention, as a key point of interest in many different research fields. In principle, events occurring during a surgery operation can be inherently life threatening. Today, the information and communications technology provide a better control and monitoring of the environment: operating rooms equipped with Ambient Intelligence techniques \cite{6579688} and smart sensing are capable of providing an higher level of awareness during most delicate procedures.
The operating room during a surgery operation can be thought as a complex environment where medical staff and patients, with their reciprocal interactions, determine the outcome of medical tasks and consequently the health conditions of the patient. Advanced sensor technologies and Artificial Intelligence techniques for data processing allow to build up systems capable of modeling and tracking the dynamics in the operating room, in the same way as the Flight Data Recorder is used to track vital information about the aircraft during flights, in order to find the causes of accidents such as equipment failures, human errors, bad weather conditions, etc.
Variety of data such as audio, video and biomedical signals can be collected in the operating room during surgery operations. They provide a valuable source for identifying errors and events that occur during complicated surgical tasks, like laparoscopy operations. A suitable utilization of the data collected during the operation can provide valuable material and information. This can be used for training the novice medics and staff, or for improving more experienced personnel, by reviewing past errors and mistakes, in order to improve the efficiency and quality of service in the operating room. 

Detecting errors by reviewing data is a complex task which requires lots of focus and a high degree of expertise. Due to the length of the operations and the limited (often cluttered) view of the environment, it is easy to miss some errors or events during the reviewing procedure. Therefore, there is a need to implement a system for assisting the reviewers in detecting and categorizing errors, to process the data collected in a meaningful way and to analyze them in order to identify relevant patterns and regularities.
 
Machine learning algorithms and Soft Computing techniques have been applied in many heterogeneous fields to recognize hidden patterns within rich and massive database, to classify structured or unstructured data represented with complex data structures \cite{bianchi2014granular, livi2013dissimilarity, bianchi2013matching}, to perform reliable predictions \cite{Bianchi2015204} and to provide smart data-driven decisions. Recently, several of these techniques have been used for improving medical procedures. Data analysis methodologies have been successfully used in the healthcare environment \cite{holzinger2014knowledge}. In particular, for what concerns data mining applications, one of the most interesting field of study is the analysis of big data in the healthcare environment. Personalized medicine, disease modeling, medical training, data-driven patient and provider education are few of the many existing applications.

With this position paper we propose a framework of an expert system for supporting decisions related to surgical tasks and operations, implemented with a layered structure. Data are initially processed by the first layer and they are transferred to the next layer. Characterizing features are successively extracted and then, in the last layer, more complex operations are implemented like errors classification procedures and automated video tagging routines.

The remainder of the paper is organized as follows. In Sect. \ref{sec:dataDesc} we describe the typology of data we focus on for our applications. In Sect. \ref{sect:sysOver} we present an overview of the structure of the system, while in Sects. \ref{sec:dataElab}, \ref{sec:patternRec}, \ref{sec:decisionLayer} we illustrate the functionality of each layer and the implemented procedures. Finally, in Sect. \ref{sec:concl} we discuss our conclusions.


\section{Description of the Data}
\label{sec:dataDesc}
The data sets that we consider in this work, which can be processed with a series of different techniques, concern:

\begin{itemize}
	\item[--] a collection of errors and events gathered in the operating room during surgical operations,
	\item[--] video recorded during operations which are associated to an audio channel,
	\item[--] kinematic data relative to movements and displacement of the personnel in the operating room,
	\item[--] biomedical signals.
\end{itemize}

Errors and events have multiple definitions in healthcare and in the surgical field, depending on the area of research and on the specific medical groups involved. The World Health Organization guidelines defines an event as ``Any deviation from usual medical care that causes an injury to the patient or poses a risk of harm" while an error is "the failure of planned actions to achieve their desired goal" \cite{reason_understanding_1995}. Another possible definition for error can be ``an act, assertion, or belief that unintentionally deviates from what is correct, right, or true and disrespect of basic surgical principles" \cite{BJSBJS9168}.
In general, an error is strictly tied to factors which lead to adverse outcomes; an exhaustive discussion about technical errors in laparoscopic surgery can be found in \cite{bonrath2013defining}.
However, from a practical point of view, the definition for error varies greatly, making a comparison of error rates between groups impractical. Complexity of scale design and subjectivity in ratings have resulted in limited use of these scores outside the experimental setting. 
A powerful and effective framework which has been defined in \cite{bonrath2013error} is the Generic Error Rating Tool (GERT) designed for surgical operations. The framework provides the definition of four different types of errors: inadequate use of force or distance (too much); inadequate use of force or distance (too little); inadequate visualization; and wrong orientation of instrument or dissection plane. Errors are ranked according to their dangerousness, with a score from 1 to 5 (5 is the most dangerous error that could lead to death). Errors can be identified during 9 different surgical tasks (abdominal access; use of retractors; use of energy devices; grasping and dissection; cutting, transection and stapling; clipping; suturing; use of suction; and other unclassified).

The video and audio data are gathered from different sources and then they are successively combined and synchronized during a preprocessing procedure. The video data are collected using different recording units: an internal camera, which is inserted in the body of the patient during laparoscopic operation, and external cameras which monitors the whole operating room.
The cameras operate at 5 fps with a frame amplitude of $320 \times 240$ pixels. The audio data comes from a single stream. The audio data can be used to identify changing phase of surgical process and to discover speech pattern and sharp sound in order to flag an event.
 
For what concerns data relative to movement and position, kinematic data are captured using a Microsoft Kinect depth-sensing device \cite{zhang2012microsoft}. Such data are generally used for performing motion analysis, for obtaining quantitative measures of the displacements of the participants in the room, for activity recognition, clustering and other performance-impacting procedures \cite{dutta2012evaluation, chen2013survey, pedro2012kinect}.

In order to enforce security of the data, cryptography assures secure data aggregation \cite{ahmad2013security}. The access control can be done by assigning the authorized user a "key" and only people who have the key can access to the data \cite{amiri2014optimizing}. Anonymity of the face of people in the surgery room deals with some multimedia techniques handling video compositing and some special effects which  blur or pixelate part of a video image \cite{chi2014facial}. Voice changer or voice enhancer is a system to alter a person's voice.  Voice changer modifies the pitch or tone, or add distortion to the voice \cite{zhou2015systems}. Also, a combination of all these techniques can be used to make someone's voice unrecognizable, in order to conceal the identity of the person in the video.


\section{System Overview}
\label{sect:sysOver}
\begin{figure}[ht!]
 \centering
 \includegraphics[scale=0.9]{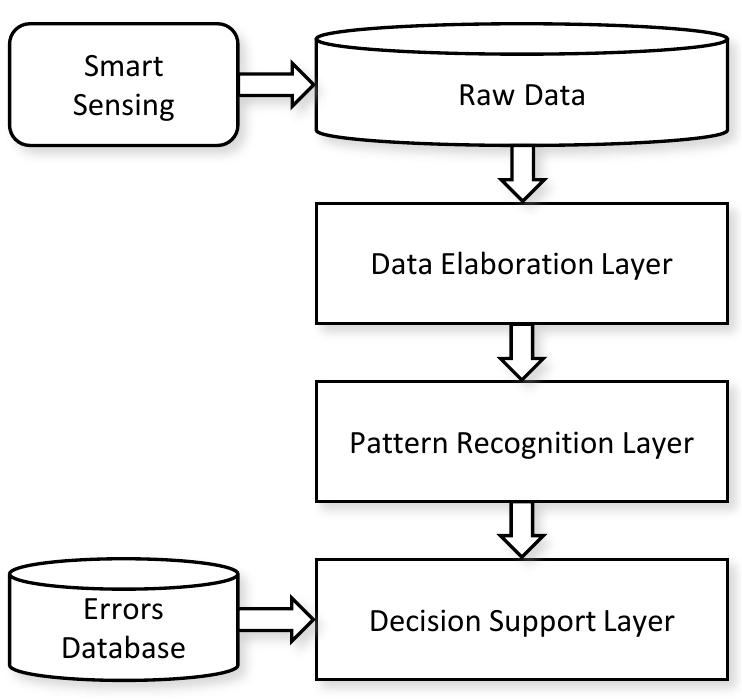}
 \caption{Overview of the whole system. Data are collected from sensors, like camera, microphone and the Microsoft Kinect. In the Data Elaboration layer, the features are extracted from the raw data and they are represented with suitable data structures. In the Pattern Recognition layer, the preprocessed data are analyzed in order to retrieve meaningful features and patterns. Finally, in the decision support layer, complex procedures like errors classification or video tagging are performed. We also use the dataset of collected errors during the training procedure of data-driven models.}
 \label{fig:sysOver}
\end{figure}

We propose a system for deriving complex decisions, starting from preemptive analysis on low-level data. The system is organized in different layers: each layer receives as input the results returned by the previous level and it produces an output with higher semantic complexity. An overview of the system is presented in Fig. \ref{fig:sysOver}. The audio, video and kinematic data (which have been previously described in Sect. \ref{sec:dataDesc}) are collected from smart sensors and they are initially processed in the first layer, that is named \textit{Data Elaboration Layer}, where different procedures can be applied for extracting numeric features from the raw data. Once the features have been generated, they are presented as input to the next layer, the \textit{Pattern Recognition Layer}, which tries to identify relevant patterns and regularities in the embedded features space. Finally in the last layer, which is called \textit{Decision Support Layer}, expert systems are used for making high level decisions, which represent the final output of the overall system.

In the following sections we describe more in depth each layer and we present few different methodologies, which can be used for implementing the functionality characterizing each layer.

\section{Data Elaboration Layer}
\label{sec:dataElab}

This layer is responsible of processing raw-level information contained in the source data and to extract numeric features, that can be used by pattern recognition algorithms. Various techniques are proposed in the literature of machine learning and signal processing for this kind of elaboration. In our case, we consider video, audio and kinematic data and in the following we review examples of useful methodologies which can be found in the literature for processing these families of databases.

The audio that is recorded in the operating suite provides relevant information about an ongoing surgery procedure. In fact, the audio stream coming from the operating room contains medical staff voices, environmental noise and, in general, sounds relative to the surgery operation. For example, the sound of the crashes of medical instruments has a characteristic signature in the Fourier spectrum \cite{5625986}. Through the application of standard filtering technique combined with more advanced signal processing approaches, such as wavelet transform or adaptive filtering techniques, meaningful features can be extracted. Moreover, as concerns the raw data pre-elaboration task, dimensionality reduction techniques such as the Principal Component Analysis (PCA) \cite{jolliffe2002principal, 7286732} or other ancillary filtering techniques can be applied for noise cancellation or for reducing the dimensionality of the data, removing less relevant components. With a suitable synchronization of the audio and the video channels, such features can be correlated to events that can lead to errors and they can turn out to be useful in the procedure of video tagging and error identification. The audio components can be added to other exogenous variables extracted from different sources, with the aim of building up complex and structured patterns to be analyzed with data mining techniques.

For what concerns processing of biomedical signals a plethora of techniques can be adopted depending on the dynamic behavior of the underlying system. Besides standard methods deriving from Fourier spectral analysis, other procedures can be adopted for extracting useful features to characterize the ongoing surgery operations. This is the case of Empirical Mode Decomposition (EMD) and fractal analysis, that are gaining an increasing attention in academia, especially in the presence of high-varying, non-stationary and non-linear signals deriving from the underlying dynamic of biological systems, ruled by non-linear equations. 

The core procedure in EMD is based on the Hilbert–Huang transform (HHT) method, which decompose nonstationary time-series originating from non-linear systems in an adaptive fashion without predefined basis function \cite{karagiannis2011noise}. In fact, the most appealing feature in EMD is the using of data-driven procedure for discovering suitable basis functions. EMD has been used in many biomedical applications such as: ECG enhancement artifact \cite{blanco2008ecg}, heart rate variability (HRV) detection \cite{echeverria2001application}, R-peak detection \cite{nimunkar2007r}, analysis of respiratory mechanomyographic signals \cite{torres2007application} and tracking of human heartbeat \cite{yeh2010intrinsic}.

Several methods have been proposed to extract and to analyze the fluctuations of the stationary component of a time-series, like detrended fluctuation analysis (DFA), detrended moving average, wavelet leaders, adaptive fractal analysis \cite{riley2012tutorial} and the so-called geometric-based approaches.
The DFA has been generalized in the so-called Multifractal Detrended Fluctuation Analysis (MFDFA) \cite{kantelhardt2002multifractal,PhysRevE.74.016103,bashan2008comparison}, a tool for analyzing time-series that describe long-term memory processes. MFDFA accounts for multi-scaling, that is, different scaling behaviors on different portions of data, which are thus identified by different sets of scaling exponents.
Time-series that are characterized by a multifractal structure can be processed using the MDMFA procedure, which provides a set of numeric features suitable for characterizing biomedical signals.
It was studied that different typologies of biomedical signals manifest a multifractal structure in healthy conditions, which is compromised when the patient shows medical conditions. For this reason, MDMFA can be used as an useful tool for the diagnosis, which is capable to distinguish healthy and pathological situations. MDMFA has been successfully applied in the work presented in \cite{ihlen2012introduction}, where it is proposed a comparison of the multifractal analysis of biomedical signals such as heartbeat of the patient before and during the surgery. 
The signal should be recorded before the beginning of the operation and for a certain period of time, in order to generate a specific pattern which characterizes the patient. In particular, multifractal coefficients such as the Hurst exponent and the multifractal spectrum are evaluated and recorded.  During the surgery, the biomedical signals are analyzed in real-time to see whether there is a significant change in multifractal coefficients. Any relevant modification in the mentioned factors during the surgery, could be linked to an emergency situation that could be the consequence of an error or of an event.


\section{Pattern Recognition Layer}
\label{sec:patternRec}
The second layer serves as the recognizing system for tracking complex actions and behaviors. One of the main objectives of data mining and knowledge discovery techniques is the retrieval of relevant features characterizing the data \cite{bianchi2014interpretable} and the identification of regularities, which in many cases are hidden and they cannot be discovered by human inspection or with basic instruments for analysis. 
Non-trivial correlations can exist, for example, between distinct types of error, among the different phases of the surgical procedure where the errors have been identified, or in the sequence of values (collected in a time-series) generated from biomedical signals. An important aspect in a knowledge discovery procedure is the definition of a suitable distance function, used for evaluating the dissimilarity of the elements in the database. Such dissimilarity measure must be configured properly in order to catch the most important and relevant patterns in the dataset. 

\subsection{A technique for knowledge and cluster discovery}
\label{method:knowledgeDiscovery}
Cluster analysis is a method used for identifying groups of similar objects represented in both geometric and non-geometric spaces \cite{bianchi2014two}. The outcome of the procedure strongly depends on the dissimilarity measure employed and on the choice of the parameters used for its tuning. Determining \textit{a-priori} the correct configuration of the distance function is an hard task that requires an amount of knowledge, which is many case is not available. A recently developed data-mining methodology presented in \cite{bianchi2015agent}, tries to automatically identify recurrent patterns and regularities among the data, by considering different sets of features which are not known in advance. The system is cluster-based and it can operates on generic type of data (vectors, labeled graphs, labeled sequences etc..), once a suitable dissimilarity measure is defined. A standard clustering algorithm generates a partition of the dataset by assigning each data element to a specific group (cluster). The clustering procedure tries to maximize both the similarity of the elements contained in a given cluster and the dissimilarity between elements pertaining to different clusters. 

In the approach proposed in \cite{bianchi2015agent}, the same element can be assigned to different clusters, by considering different sets of relevant features, which are automatically discerned by the algorithm. This can be done automatically by tuning the configuration of the parameters within the employed dissimilarity measure, so that it is possible to generate clusters which are compact, sufficiently populated and separated from the remaining part of the dataset, using different metrics for evaluating their similarity.
The output of the procedure is a set of well-defined clusters and the configurations of the dissimilarity measure that have been used for identifying each cluster. Such configurations of the parameters can be used in a semantic interpretation of the results, in order to characterizes a cluster in terms of the features for which the elements contained inside result to be similar.

In general, the clusters produced with this precedure do not form a partition. In fact, one element can belong to different clusters at the same time, since it can be considered similar to different groups of elements, depending on which set of features are being considered relevant in the dissimilarity measure. Furthermore, if an element cannot be associated to any cluster which is sufficiently compact, populated and separated from the remaining elements of the data, it will not appear in the final output and it will be marked as an \textit{outlier}.
Finally, it is possible that some clusters are very similar in terms of the elements contained, but they have been identified considering different sets of features. In this case, such features are classified as \textit{equivalent} with respect to the elements contained in the cluster under consideration.
The results can be used in a semantic characterization and in a qualitative analysis on the considered data; this results particularly effective in dataset where there are no information available on the nature of data. The relevant clusters which are returned can be used for synthesizing a cluster-based classification system, that is capable of identifying and characterizing different types of errors like the ones pertaining the surgical operations.

\subsection{Actions and activity recognition}
\label{method:activityRec}
Advanced video acquiring devices that are capable of collecting depth data coming from depth sensors, such as the Microsoft Kinect, can be used together with machine learning algorithms for tracking activities in the operating suite and for relating events to the errors occurring during surgery operations. The advance in technology in building smaller acquiring devices allows their usage in delicate environments, like healthcare facilities. Moreover, well-known computer vision libraries, together with suitable designed algorithms, allow to track multiple persons in a room and to register their paths and actions exploiting depth imagery \cite{Chen20131995}. Once measured the amount of movements in the room and the interactions "between who", it is possible to determine whether or not the tracked interactions follow a characteristic pattern. The last is an important point because some surgery operations follow a strict protocol and any deviation from it, can be interpreted as an alarm for an imminent dangerous event or an error.
 
Once high level schemes of the time-sampled actions and behaviors have been generated using procedures and algorithms from the time-series analysis framework and graph theory, it is possible to establish a causal relationship among the moving patterns of the medical staff. Such displacements can be correlated with the multitude of events that occur in the operating room. Hidden Markov Models, the generative probabilistic models used for generating hidden states from observable data, are powerful candidates for activity recognition \cite{4475859,7096216}. Additional frameworks such as Neural Networks and Fuzzy Logic (FL) can be used in these challenging tasks. 

\section{Decision Support Layer}
\label{sec:decisionLayer}

In the last layer, expert systems are employed for performing high-level semantic deductions, according to the patterns and the regularities which have been identified in the previous layers. In particular, here we focus on the problem of video tagging and errors classification.

\subsection{Automatic assistance in the process of errors discovery}
The process of error discovery in surgery application is a challenging task. After their recordings, surgery videos are reviewed by experienced doctors in order to find and to tag errors occurred during the procedure. Correctly identified errors, which come with a proper description and an explanation on how they can be avoided, can be used to enhance the quality of service in surgical, they represent a method for controlling the performances of the personnel and for training the medical staff. The procedure of tagging the video with identified errors requires a lot of focus, time and technical preparation. Exogenous sources of information and automation procedures can be used as a support for this task. 

A possible approach to video tagging is based on FL techniques, used to build up the expert systems.
FL, a multi-valued logic introduced in the engineering  field by Lotfi Zadeh, can be used effectively in applications dealing with uncertainty in the data. Fuzzy sets are known for their capability of modeling notions of continuity into deductive thinking \cite{Fuzzy_anesthesia}. Practically, this means that fuzzy sets allow the use of conventional symbolic systems, in the form of tabulated rules, in continuous form. Thus, FL allows uncertainty reasoning mechanisms starting from uncertain and noisy data and it uses a set of rules that can be expressed in a human-like form \cite{6608437}. The power of FL reaches his maximum strength in the expert systems field where it is possible to build up a system that can take decisions autonomously, by exploiting expert knowledge encoded in a computer system together with an inference engine.   

\begin{figure*}[ht!]
 \centering
 \includegraphics[scale=0.6]{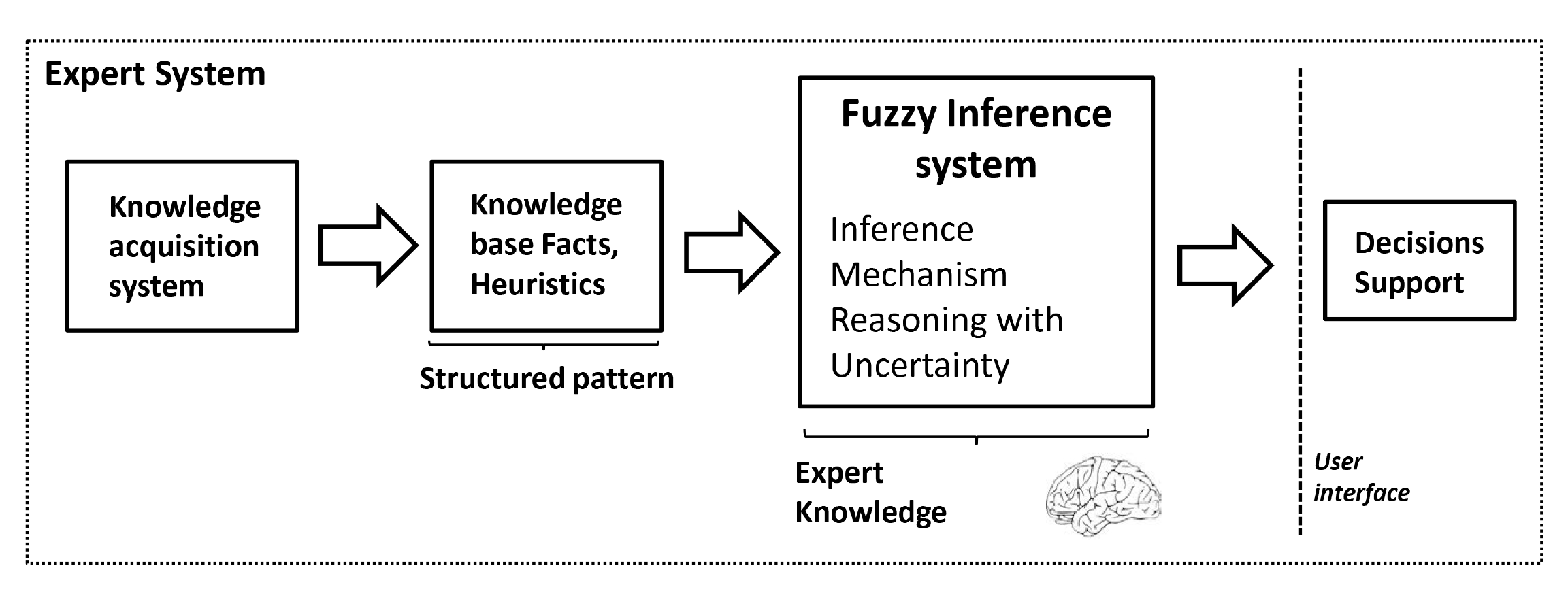}
 \caption{Functional scheme of the proposed Expert System for Decision Support in surgery errors discovery.}
 \label{fig:Expert_S}
\end{figure*}

Nowadays, the fuzzy technology is widely adopted in medicine and healthcare for dealing with missing values in database, impreciseness and sometimes contradictory nature of the data \cite{Abbod2001331}. In particular, using FL-based methodologies is really effective when the processes are too complex for being analyzed with conventional quantitative techniques or when the available sources of information are interpreted qualitatively, inexactly, or uncertainly. Therefore, an expert system that performs fuzzy reasoning mechanisms can be designed to create a Decision Support System (DSS) based on a knowledge system,  i.e. a repository of problem domain knowledge embodied in the DSS as either data or procedures, to tackle the problem of assisting the discovery of errors and events in the operating room. Fuzzy technology is designed to address the problem of data fusion, which allows to aggregate data that can differ not only from the syntactic point of view, but also from the semantic one. 
As depicted in Fig. \ref{fig:Expert_S}, the proposed core module of the Expert System for decision support is the inference engine that performs approximate reasoning by means of simple rules defined by experts (which are the surgeons in our case). More precisely, the inference engine is fed with structured objects returned by the Data Elaboration and the Pattern Recognition layer, which are the features extracted from audio, video or activity stream, that are suitably transformed in fuzzy elements. The output of the system, after applying the fuzzy inference rules and the de-fuzzification operator, is a degree of confidence for a specific error type in the operating suite. 

Since a FL system provides a mapping from a set of input variables to a set of output variables, a Multiple Input Multiple Output (MIMO) fuzzy inference system can be synthesized. 
The MIMO system maps features extracted from audio sources, such as the spectral energy in a predetermined frequency bands, together with features extracted from different channels, like video stream or kinematic data, to the various class of errors and their respective scale of severeness, according to the error model employed (see GERT in Sect. \ref{sec:dataDesc}). 
Relative to video sources we can evaluate the compression factor in a predetermined time interval \cite{5625986}, while trough depth sensors we can analyze the path generated during the tracking. 
Taking into account the possibility of marking the features extracted (input) with a time-stamp which synchronize their occurrence within a video of an operation (for example, a frame which depicts the organ concerned in the video), the output information can provide an automatic warning about a possible occurrence of errors. 
In other words, the inference system, which exploits other heterogeneous information captured in the operating room environment, alert the user which is reviewing the video, through a suitable software interface, the possible occurrence of a particular type of error. Finally, such knowledge can be used to highlight some parts of the video, easing and improving the video tagging task performed by expert doctors.

\subsection{Error Classification}
Although a universal definition and categorization of surgical errors does not exist, several attempts have been made to define classes of errors. In \cite{bonrath2013defining} a method is proposed to cluster and categorize different types of technical errors according to some characterizing features.
A widely used and effective approach, applies clustering techniques for identifying similar patterns among the data. Such clusters are used to define different classes of error according to the values contained and to the adopted concept of similarity.
While many approaches rely on a crisp logic, the fuzzy clustering implements a soft clustering schema. Each member of the cluster belongs to that group with a degree of membership. Therefore, there is no solid boundary between clusters \cite{bezdek1984fcm}. IN our applicative context, an algorithm like the Fuzzy C-mean (FCM) assign no hard boundary to different types of errors in operation room and one type of error might be a member of two distinct groups. 

Errors during surgery can be defined as a sequence of actions that deviate from a specific, predefined protocol. Hence, errors are \textit{novelty patterns} that differ from standard patterns. Novelty detection solutions based on Support Vector Data Description (SVDD) \cite{SVDD_TAX} can be adopted in order to discover error patterns in a stream of data. In the last case, through the one-class classification framework \cite{one-class_survey__2010,6889668}, a decision region can be constructed upon discovered clusters to train the model on available target or standard patterns. On the basis of a parametric dissimilarity measures, whose weights are assigned to each feature and they are learned through an optimization technique and a suitable decision region (DR), error patterns can be classified and recognized. Once the dissimilarity parameters and the decision region model are obtained, a simple Boolean decision rule can be adopted depending on the distance/dissimilarity of a given element from the representative of the decision region. If a given pattern, coming from the Elaboration layer fed by smart sensors in the surgery room, falls into the DR, then it is classified as a standard or a target pattern. Dealing with complex patterns, constructed with heterogeneous information, makes it difficult to define a precise boundary between an error and a standard surgery procedure. Moreover, fuzzy techniques can be used to assign a suitable score to the Boolean classifier in order to measure, in a continuous fashion, the reliability of the decision, providing a soft discrimination between patterns that belong to different classes \cite{DeSantis2015}.


\section{Discussion}
\label{sec:discussion}

At first glance, the framework that we describe for transforming the operating room in a cognitive environment with an high degree of awareness, could seem a complicated and huge system to build. However, thanks to the high level of modularity in the structure proposed, it is possible to divide the implementation of the system, which can be carried on in different phases, by different teams which use heterogeneous instruments and competences, as long as an agreement in the communication protocols of the API in each layer is guaranteed. Whenever a module become obsolete or a new technology is available for a possible improvement, new features can be added to each layer and it can be updated independently, without modifying the remaining part of the architecture.

If we consider the direction in which the technology evolution is moving, given the decrement of the prices in the hardware required for smart sensing, the increment of the performances of ubiquitous computing technologies, the recent progresses of the research on Ambient Intelligence and the efficiency of machine learning algorithms, we realize that the creation of a smart operating theater is just a matter of time. 
As in the case of flight data recorder technology \cite{ratchford1984digital}, the possibility of tracking and elaborating an increasing volume of data, allows to process additional data sources coming from more sophisticated sensors and equipments on board. 

A downside of the proposed framework, is the difficulty in implementing a real-time monitoring system due to the amount of data to be analyzed and the computational complexity of the learning algorithms used for processing the information. 
Modern frameworks can be adopted for dealing with the computational complexity, which can reduce the running time of demanding algorithms by using parallel computing architectures, like multi-core CPU, GPU programming, multi-threading and distributed computing. 
The choice of a particular architecture depends on: i) the trade-off between the cost of the resources (in terms of time, hardware and space of occupancy) and the performance required ii) an understanding of which routines and sub-routines in algorithms can be implemented with a parallel architecture, along with a definition of a suitable measure for evaluating the speedup of the possible implementation strategies.  

Nowadays, well-established frameworks for managing Big Data applications, such as MapReduce and the Hadoop platform, can be adopted for dealing with applications that demands a large quantity of memory for elaborating raw data.

It is well know that in healthcare facilities and especially in the surgery rooms, there are a set of strict rules and well-defined protocols, which limit the freedom of action, the flexibility in applying certain kinds of techniques and the use of some instruments. In fact, an actual issue is the invasiveness of additional hardware in the smart operating room. However, this does not represent a problem since the framework proposed uses a minimal amount of sensors in the operating suite and it does not require the use of ``markers" for tracking medical staff or surgical instruments. 

In fact, many of the instruments that have been considered for retrieving data are placed away from the operating theater. This facilitate, through of the proactive collaboration with medical experts, to deal with hygienic issues, that can impair the asepticity of the operating environment.


\section{Conclusions}
\label{sec:concl}
The ``injection of intelligence" in the operating room is a challenging task, which is required in the process of transforming hospital facilities in "Smart Hospitals". This position paper focuses on an interdisciplinary, community-based approach designed to complement existing healthcare services, specifically the surgery operations, with machine learning and pattern recognition techniques. In particular, by thinking at the surgery room as a complex environment with its own dynamic generated from reciprocal interactions between medical staff and patients, it is possible to equip the operating room with smart sensors and suitable hardware. The data generated are successively processed with machine learning algorithms, in order to facilitate decision making and to provide support during and after the surgery. 

Techniques such as signal processing, data mining, pattern recognition and classification, can be integrated in a unique intelligent system that is capable to extract meaningful and semantically rich information, which are useful for understanding and also for discovering errors and events that can occur during surgery operations. 
The last task is essential for training medical staff and the proposed research goes in this direction. Raw data coming from audio, video, and depth sensors, after a pre-elaboration and a layered processing through machine learning algorithms, are transformed in complex patterns that are fed to an expert system for supporting the decisions taken, for example, in the video tagging procedures during the review process performed by medical experts. Moreover, the proposed methodology natively facilitates the integration of qualitative knowledge of experts in automatic computational frameworks. Data analysis and system design involve the active collaboration with experienced partners, which provide recommendations on how to interpret research results, in order to achieve the desired goals. The preliminary research stage based on data acquisition and analysis, as well as the realization of the system, is able to provide a huge dissemination of results helping the final users. By developing a user-oriented cognitive and dynamic health-care environment, we are confident that the final product can take part of the operating room environmental design, where new methodologies can be practiced in order to better understand the outcome of an operation and the nature of surgery errors and events.

All the methodologies that have been discussed in this position paper are well consolidated and several stable software implementations, which are provided by many software libraries, are available for different programming languages. This allows a simple and rapid implementation of the features in the system that we propose, as well as an easy integration with the devices, the instruments and the procedures for data collection used in the operating room environment.

\bibliographystyle{abbrvnat}
{\footnotesize
\bibliography{Bibliography}}


\end{document}